\documentclass[aps,prd,preprint,groupedaddress]{revtex4-1}
\usepackage[utf8]{inputenc}
\usepackage[english]{babel}

\usepackage{amsmath,amssymb,amsfonts,enumerate,float,graphicx,hyperref,mathrsfs,mathtools,fancyvrb,slashed,pdfpages,empheq,tabularx,caption,multirow,subcaption}

\usepackage[inline,shortlabels]{enumitem}
\allowdisplaybreaks

\graphicspath{{./figures/}}

\newcommand{\pp}[1]{\left ( #1 \right )}
\newcommand{\bb}[1]{\left [ #1 \right ]}

\newcommand{\bra}[1]{\left\langle #1 \right\vert}
\newcommand{\ket}[1]{\left\vert #1 \right\rangle}

\DeclareMathOperator{\tr}{\mathrm{tr}}

\newcommand{\e}[1]{\mathrm{e}^{#1}}

\newcommand{\nn}{\nonumber\\ &}

\makeatletter
\AtBeginDocument{\let\LS@rot\@undefined}
\makeatother

\def\el{\varepsilon}
\def\elx{\varepsilon \cdot x}
\def\qx{q \cdot x}
\def\gper#1#2{g^\perp_{#1#2}}

\def\V{\mathcal V (\alpha_1,\alpha_3)}
\def\A{\mathcal A (\alpha_1,\alpha_3)}

\def\Iu{\int_0^1 \d u\ \e^{\I \bar u \qx}}
\def\Ia{\int \mathcal D \vec \alpha \ \e^{\I (\alpha_1 + u \alpha_3) \qx}}

\def\d{\mathop{}\!\mathrm{d}}
\def\D#1{\mathop{}\!\mathrm{d^#1}}
\def\e{\mathrm{e}}
\def\I{\mathrm{i}}

\DeclareMathOperator\IA{\mathcal I_\alpha}

\begin{document}
	\title{Strong coupling constants of charmed and bottom mesons with light vector mesons in QCD sum rules}
	
	\author{T. M. Aliev}
	\email[]{taliev@metu.edu.tr}
	\affiliation{Physics Department, Middle East Technical University, Ankara 06800, Turkey}
	
	\author{K. \c Sim\c sek}
	\email[]{ksimsek@u.northwestern.edu}
	\affiliation{Department of Physics \& Astronomy, Northwestern University, Evanston, Illinois 60208, USA}
	
	\date{\today}
	
	\begin{abstract}
		We estimate the strong coupling constants of charmed and bottom mesons $ D_{(s)}^* $, $ D_{(s)1} $, $ B_{(s)}^* $, and $ B_{(s)1} $ with light vector mesons $ \rho $, $ \omega $, $ K^* $, and $ \phi $ within the framework of light-cone QCD sum rules. We  compare our estimations to the ones predicted by other approaches.
	\end{abstract}
	\maketitle 
	\section{Introduction}\label{sec:1}
	Strong coupling constants between heavy and light mesons are among essential ingredients for the description of low-energy hadron interactions. Precise determination of these couplings can provide key information for studying the nature of heavy mesons. In particular, they stand as a useful source in the investigation of final-state interactions of $ D $ and $ B $ meson decays. Moreover, for understanding the production and absorption cross sections of the $ J/\psi$ meson in heavy-ion collisions, vertices involving charmed mesons are needed. On the other hand, heavy-light meson couplings are fundamental objects since they carry information about the low-energy behavior of QCD. However, this region is far away from the perturbative region of QCD. Therefore, for a reliable estimation of these couplings, some nonperturbative approach is required. 
	\par 
	Among the nonperturbative approaches, the QCD sum rules method \cite{Shifman1979} occupies a special place. This method is based on the fundamental QCD Lagrangian and includes nonperturbative effects. 
	\par
	So far, to the best of our knowledge, many heavy-light meson vertices such as 
	$ D_s^*D^*K $, $ D_{s1}D_1K^* $ \cite{Janbazi2018}, 
	$ D^*D^*\rho $ \cite{Bracco2008}, 
	$ D^*D\pi $, $ B^*B\pi $ \cite{Navarra2000, Navarra2001}, 
	$ DD\rho $ \cite{Bracco2001}, 
	$ D^*D\rho $ \cite{Rodrigues2011}, 
	$ DDJ/\psi $ \cite{Matheus2002}, 
	$ D^*DJ/\psi $ \cite{Silva2004}, 
	$ D^*D^*\pi $ \cite{Carvalho2005}, 
	$ D_sD^*K $, $ D_s^*DK $ \cite{Bracco2006}, 
	$ DD\omega $ \cite{Holanda2007}, 
	$ D_{s1}D^*K $, $ D_{s1}D^*K_0^* $ \cite{Gamermann2007, Ghahramany2012}, 
	$ D_sD_sV $, $ D_s^*D_s^*V $, $ D_{s0}^*D_{s1}V $, $ D_sD_s^*V $ \cite{Khosravi2013, Khosravi2014}, 
	$ D_1D^* \pi $, $ D_1D_0\pi $, $ D_1D_1\pi $, $ B_1B^*\pi $, $ B_1B_0\pi $, $ B_1B_1\pi $ \cite{Janbazi2014}, 
	$ D^*D^*J/\psi $ \cite{Bracco2005}, 
	$ B_{s0}BK $ \cite{Bracco2010}, 
	$ B_s^*BK $ \cite{Cerqueira2012}, 
	$ D_s^*D_s\phi $ \cite{Yu2015}, 
	$ D_sDK_0^* $, $ B_sBK_0^* $, $ D_s^*DK $, $ B_s^*BK $, $ D_s^*DK_1 $, $ B_s^*BK_1 $ \cite{Sundu2011}, 
	$ D^*D\pi $, $ D^*D^*\pi $, $ DD\rho $, $ D^*D\rho $, $ D^*D^*\rho $, $ DDJ/\psi $, $ D^*DJ/\psi $, $ D^*D^*J/\psi $ \cite{Bracco2012},
	$ D_sDK^* $, $ D_sD^*K^* $ \cite{Janbazi2018b},
	$ D_s^*DK^* $, $ B_s^*BK^* $ \cite{Azizi2011},
	$ D_0D_{s0}^*K^* $, $ D_1D_{s0}^*K^* $ \cite{Asgarian2020},
	$ B_sB^*K $, $ B_sBK^* $ \cite{Cerqueira2015},
	$ B^*B^*\rho $ \cite{Cui2012},
	$ B_{s1}B^*K $, $ B_{s1}B^*K_0^* $ \cite{Ghahramany2012b},
	$ B_{s1}^*B^*K $ \cite{Cui2012b},
	$ D_sD_sJ/\psi $, $ D_sD_s\phi $ \cite{Bracco2012b},
	$ D^*D_s^*K $, $ D_1D_{s1}K $, $ D^*D_sK $, $ D_1D_{s0}^*K $ \cite{Seyedhabashi2020}, and
	$ B_{s0}B_1K $, $ B_{s1}B_1K $ \cite{Asgarian2021}
	are estimated in the framework of the three-point QCD sum rules (3PSR),
	$ D^*D_sK $, $ D_s^*DK $, $ D_{s0}DK $, $ D_0D_sK $ \cite{Wang2006}, 
	$ D^*D^*P $, $ D^*DV $, $ DDV $ \cite{Wang2007},
	$ D^*D\rho $, $ B^*B\rho $ \cite{Wang2020},
	$ D^*D\pi $, $ B^*B\pi $ \cite{Belyaev1995}, and
	$ D^*D^*\rho $ \cite{Wang2008}
	with the light-cone QCD sum rules (LCSR) approach,
	$ DD\rho $, $ D^*D^*\rho $ \cite{Bayona2017} in holographic QCD (HQCD),
	$ DD\rho $, $ D^*D\rho $, $ D^*D^*\rho $ \cite{ElBennich2017} using the Dyson-Schwinger equation (DSE) in QCD,
	$ D^*D\pi $, $ DD\rho $, $ D^*D^*\rho $ \cite{Abada2002, Becirevic2009, Becirevic2013, Can2013} in lattice QCD,
	$ DD\rho $ \cite{Fontoura2017} in the non-relativistic quark model,
	$ DD\rho $, $ DD\sigma $, $ D^*D^*\rho $, $ D^*D^*\sigma $, $ BB\rho $, $ BB\sigma $, $ B^*B^*\rho $, $ B^*B^*\sigma $ \cite{Kim2020} based on the approach of correlated $ 2\pi $ exchange with the pole approximation (PA),
	$ B^*B^*\rho $ \cite{Liu2009, Liu2010} and
	$ D^*D^*\rho $ \cite{Oh2001} in the meson-exhange model (MEM), 
	$ B^*B^*\rho $, $ D^*D\pi $, $ B^*B\pi $ \cite{Lee2009, Colangelo1994} in the potential model, and, lastly,
	$ D^*D^*\rho $, $ B^*B^*\rho $ \cite{Liu2019} in the one-boson exchange (OBE) model.
	\par
	In the present work, we study the coupling constants of $ D_{(s)}^* D_{(s)}^* V $, $ D_{(s)1} D_{(s)1} V $, $ B_{(s)}^* B_{(s)}^* V $, and $ B_{(s)1} B_{(s)1} V $ where $ V=\rho, \omega, K^*,\phi $ within the LCSR method. In this approach, the operator product expansion (OPE) is carried out near the light cone, $ x^2 \sim 0 $, and the nonperturbative effects appear in the matrix elements of nonlocal operators, which are parameterized in terms of the light-cone distribution amplitudes (DAs) of the corresponding hadrons, instead of the vacuum condensates that appear in the standard sum rules method \cite{Chernyak1984, Balitsky1989}. 
	\par
	The paper is organized as follows. In Sec. \ref{sec:2}, we derive the desired sum rules for the strong coupling constants of the said vertices. In Sec. \ref{sec:3}, we present our numerical analysis and resultant values for the aforementioned couplings. Sec. \ref{sec:4} contains our conclusion. 
	\section{Light-cone sum rules for charmed/bottom meson-light vector meson couplings}\label{sec:2}
	Let us introduce the following correlation function in order to compute the strong coupling constants of vector and axial vector charmed and bottom mesons with light vector mesons:
	\begin{align}
		\Pi _{\mu\nu} = \I \int \D4x\ \e^{\I px} \bra{V(q,s)} {\rm T} \{j _{\mu}(x) \bar j _{\nu}(0) \} \ket 0 \label{1}
	\end{align}
	where $ V(q,s) $ is a vector meson of mass $ m_V $, 4-momentum $ q $ and 4-polarization $ \el $ (but we will suppress the superscript $ s $ for the most part), and $ j_{\mu} $ indicates the interpolating current of the corresponding charmed or bottom meson. For the considered mesons, the interpolating current can be written as
	\begin{align}
		j _{\mu} (x) = \bar q(x) \Gamma _\mu Q(x) \label{2}
	\end{align}
	where $ \Gamma_\mu = \gamma_\mu\ (\gamma_\mu \gamma_5) $ for vector (axial vector) charmed and bottom mesons, $ Q $ is a heavy quark, namely $ c $ or $ b $, and $ q $ is one of the light quarks, $ u $, $ d $, or $ s $.
	\par
	In the framework of the LCSR method, one computes the correlation function in two different regions. On one side, it can be calculated in terms of hadrons. which is also known as the phenomenological part; on the other side, the calculation is carried out in the deep Euclidean domain, i.e. $ p^2 \to -\infty $ and $ (p+q)^2 \to -\infty $, using the OPE over twist, which is traditionally called the theoretical part. In order to suppress the contributions from excited states and the continuum, as well as to enhance the contribution of the ground state, a double Borel transformation is performed with respect to the variables $ -p^2 $ and $ -(p+q)^2 $.
	\par
	Let us begin our analysis of the derivation of the sum rules by focusing on the phenomenological part of the correlation function first. Inserting a complete set of intermediate states carrying the same quantum numbers as the interpolating currents and isolating the ground-state meson, one can obtain
	\begin{align}
		\Pi _{\mu\nu} = \frac{\bra 0 j_\mu \ket{\mathcal M_2(p_2)} \langle \mathcal M_2(p_2) V(q,s) \vert \mathcal M_1(p_1) \rangle \bra{\mathcal M_1(p_1)} \bar j _\nu \ket 0}{(p_2^2 - m_2^2) (p_1^2 - m_1^2)} + \cdots
	\end{align}
	where $ \mathcal M_1(p_1) $ and $ \mathcal M_2(p_2) $ are the initial and final charmed or bottom mesons of mass $ m_1 $ and $ m_2 $ and 4-momentum $ p_1 $ and $ p_2 $, respectively, and $ \cdots $ indicates the contributions from higher states. The matrix elements in this correlation function are given by
	\begin{align}
		\bra 0 j_\mu \ket{\mathcal M_2(p_2)} &= f_2 m_2 \varepsilon_{2,\mu}, \\
		\bra {\mathcal M_1(p_1)} \bar j_\nu \ket 0 &= f_1 m_1 \varepsilon_{1,\nu}^*, \\
		\langle \mathcal M_2(p_2) V(q,s) \vert \mathcal M_1(p_1) \rangle &= g (
			p_2 \cdot \varepsilon^* \varepsilon_1 \cdot \varepsilon_2^* 
			+ p_1 \cdot \varepsilon^* \varepsilon_1 \cdot \varepsilon_2^*
			+ q \cdot \varepsilon_1 \varepsilon^* \cdot \varepsilon_2^*	
			- p_2 \cdot \varepsilon_1 \varepsilon^* \cdot \varepsilon_2^*
			\nn - p_1 \cdot \varepsilon_2^* \varepsilon^* \cdot \varepsilon_1^*
			- q \cdot \varepsilon_2^* \varepsilon^* \cdot \varepsilon_1
		)
	\end{align}
	where $ q:= p_2 - p_1 $ is the transfer 4-momentum. From now on, we will let $ p := p_1 $ and use $ q $ instead of $ p_2 $ through $ p_2 = p+q $. Making use of the spin sum over the 4-polarization vectors $ \varepsilon_1 $ and $ \varepsilon_2 $ given by
	\begin{align}
		\sum _s \varepsilon_{\mu}^s \varepsilon_{\nu}^{s*} = -g_{\mu\nu} + \frac{p_{\mu} p_{\nu}}{m^2},
	\end{align}
	we obtain the phenomenological side of the correlation function to be
	\begin{align}
		\Pi _{\mu\nu} = \frac{f_1 f_2 m_1 m_2 g }{[(p+q)^2 - m_2^2] (p^2 - m_1^2)} p \cdot \varepsilon^* g_{\mu\nu} + \cdots
	\end{align}
	where $ \cdots $ denotes other structures.
	\par
	In order to derive the LCSR for the strong coupling constant $ g $, one needs to compute the theoretical side of the correlation function. Afterwards, selecting the same structure, i.e. $ p \cdot \varepsilon^* g_{\mu\nu} $, and matching it to the result obtained from the phenomenological part, one arrives at the desired LCSR. We get the said part of the correlation function by making use of the OPE in the deep Euclidean region, $ p^2 \to -\infty $ and $ (p+q)^2 \to -\infty $. Inserting the interpolating currents given in the form of Eq. \eqref{2} and applying the Wick theorem to the correlation function given by Eq. \eqref{1}, we obtain
	\begin{align}
		\Pi _{\mu\nu} = \I \int \D4x\ \e^{\I px} \bra{V(q,s)} \bar q_1(x) \gamma_\mu (\gamma_\mu \gamma_5) S_Q (x) \gamma_\nu (\gamma_\nu\gamma_5) q_2(0) \ket0. \label{9}
	\end{align}
	Here, $ S_Q(x) $ is the heavy-quark propagator,
		\begin{align}
			S_Q^{aa'} (x) &= \frac{m_Q^2}{4\pi^2} (\I K_2 \slashed x + K_1) \delta^{aa'} \nn - \frac{g_s}{16\pi^2} m_Q \int_0^1 \d u\ [
				\I K_1 (\bar u \slashed x \sigma^{\lambda\tau} + u \sigma^{\lambda\tau} \slashed x) + K_0 \sigma^{\lambda\tau}
			] G^{(n)}_{\lambda\tau}(ux) \pp{\frac{\lambda^{(n)}}2}^{aa'} \label{10}
		\end{align}
	where $ G^{(n)}_{\lambda\tau} $ is the gluon field strength tensor, the $ \lambda^{(n)} $ are the Gell-Mann matrices, and we have defined the shorthand notation $ K_n := K_n(m_Q \sqrt{-x^2})/\pp{\sqrt{-x^2}}^n $, $ K_n(z) $ being the $ n^{\rm th} $ modified Bessel function of the second kind.
	\par 
	Expanding the propagator inside the correlation function and using the Fierz identities,
	\begin{align}
		q_{2\alpha}^a(0) \bar q_{1\beta}^{a'}(x) &= -\frac 1{12} \delta^{aa'} (\Gamma_i)_{\alpha\beta} [\bar q_1(x) \Gamma_i q_2(0)], \label{11} \\
		q_{2\alpha}^a(0) G_{\lambda\tau}^{(n)} \bar q_{1\beta}^{a'}(x) &= -\frac1{16} \pp{\frac{\lambda^{(n)}}2}^{aa'} (\Gamma_i)_{\alpha\beta} [\bar q_1(x) \Gamma_i G_{\lambda\tau}^{(n)} q_2(0)] \label{12}
	\end{align}
	where the $ \{\Gamma_i \}_{i=1}^5 $ is the complete set of Dirac matrices,
	\begin{align}
		\Gamma_1 = 1,\ \Gamma_2 = \gamma_5,\ \Gamma_3 = \gamma_\alpha,\ \Gamma_4 = \I \gamma_\alpha \gamma_5,\ \Gamma_5 = \frac1{\sqrt2}\sigma_{\alpha\beta},
	\end{align}
	one can see that, in the calculation of the theoretical side of the correlation function, one needs the matrix elements $ \bra{V(q,s)} \bar q_1 (x) \Gamma_i q_2 (0) \ket 0 $ and $ \bra{V(q,s)} \bar q_1 (x) \Gamma_i G_{\lambda\tau}^{(n)} q_2 (0)\ket 0 $. These matrix elements, expressed in terms of light vector meson DAs of various twists \cite{Ball1996, Ball1998, Ball1999, Ball2006}, constitute the primary nonperturbative input parameters of the LCSR. The full compilation of these matrix elements and the corresponding DAs can be found in Appendix C of \cite{Aliev2020}. For the sake of completeness, we present only the relevant expression that will appear in the computation of the correlation function in Appendix \ref{app:A} of the present work.
	\par
	At this point, we would like to make two remarks. 
	\begin{enumerate}
		[$ \bullet $]
		\item Let us consider the terms without $ G_{\lambda\tau}^{(n)} $, which conventionally provide the major contribution to the sum rules. We are interested in the structure $ p \cdot \varepsilon^* g_{\mu\nu} $. In the correlation function, there will be traces of the form $ \tr \gamma_\mu (\gamma_\mu \gamma_5) \slashed x \gamma_\nu (\gamma_\nu \gamma_5) \Gamma_i $. It turns out that only the third term in the Fierz expansion, namely $ \Gamma_3 = \gamma_\alpha $, will contribute to the aforementioned structure.
		\item We do the continuum subtraction as described in \cite{Belyaev1995}. That is to say, after taking the double Borel transform of the theoretical side of the correlation function, there will appear an exponential factor $ \e^{-m_Q^2/M^2 - m_V^2/(M_1^2 + M_2^2)} $. According to \cite{Belyaev1995}, making the replacement 
		\begin{align}
			\e^{-m_Q^2/M^2 - m_V^2/(M_1^2 + M_2^2)} \to \e^{-m_Q^2/M^2 - m_V^2/(M_1^2 + M_2^2)} - \e^{-s_0/M^2}
		\end{align}
		will suffice as far as the continuum subtraction is concerned. Since the masses of the initial and final states are the same or nearly equal, we take $ M_1^2 = M_2^2 = 2M^2 $. 
	\end{enumerate}
	Putting the expressions given in Eqs. \eqref{10}, \eqref{11}, and \eqref{12} into \eqref{9}, taking the double Borel transform over the variables $ -p^2 $ and $ -(p+q)^2 $, and using the results of Appendix \ref{app:B} to carry out the integrals encountered, we obtain the required sum rules for the strong coupling constants, $ g_V $ and $ g_A $, where the subscripts $ V $ and $ A $ stand for the case of vector and axial vector charmed or bottom mesons:
	\begin{align}
		g_V &= g_V^{(0)} + g_V^{(1)}, \\
		g_A &= g_A^{(0)} + g_A^{(1)} 
	\end{align}
	where
	\begin{align}
		g_V^{(0)} &= \frac 1{2M^6 f_1 f_2 m_1 m_2} \e^{[2(m_1^2+m_2^2) - m_{V}^2 - 4(m_Q^2+s_0)]/4M^2} \bb{\e^{(4m_Q^2 + m_{V}^2)/4M^2} - \e^{s_0/M^2}} \nn\times \bigg[
			2M^4 f_{V} m_Q^2 m_{V}^3 \hat{\hat \phi}_2^\parallel(u_0) 
			-4M^4 f_{V} m_Q^2 m_{V}^3 \hat{\hat \phi}_3^\perp (u_0)
			+2M^4 f_{V} m_Q^2 m_{V}^3 \hat{\hat \psi}_4^\parallel(u_0) 
			\nn -M^8 f_{V} m_{V} \phi_2^\parallel(u_0) 
			+M^6 f_{V} m_{V}^3 \phi_4^\parallel(u_0) 
			+M^4 f_{V} m_Q^2 m_{V}^3 \phi_4^\parallel(u_0)
		\bigg], \\
		g_V^{(1)} &= \frac1{2M^6 f_1 f_2 m_1 m_2} \e^{[2(m_1^2+m_2^2) - m_{V}^2 - 4(m_Q^2+s_0)]/4M^2} \bb{-\e^{(4m_Q^2 + m_{V}^2)/4M^2} + \e^{s_0/M^2}} \nn\times \bigg\{
			8M^2 m_Q m_{V}^6 f_{V}^T \IA[u_1^3 \hat{\hat{\hat{\mathcal T}}} (\alpha_1,\alpha_3)]
			-4m_Q^3 m_{V}^6 f_{V}^T \IA[u_1^3 \hat{\hat{\hat{\mathcal T}}} (\alpha_1,\alpha_3)]
			\nn -2M^4 m_Q m_{V}^4 f_{V}^T \IA [u_1 \hat{\mathcal T}_3^{(4)} (\alpha_1,\alpha_3) ]
			-2M^4 m_Q m_{V}^4 f_{V}^T \IA[u_1 \hat{\mathcal T}_4^{(4)} (\alpha_1,\alpha_3)]
			\nn + M^6 f_{V} m_{V}^3 \IA[\A]
			+ M^6 f_{V} m_{V}^3 \IA[(2u_1-1) \V]
		\bigg\}, \\
		g_A^{(0)} &= g_V^{(0)}, \\
		g_A^{(1)} &= \frac1{2M^6 f_1 f_2 m_1 m_2} \e^{[2(m_1^2+m_2^2) - m_{V}^2 - 4(m_Q^2+s_0)]/4M^2} \bb{-\e^{(4m_Q^2 + m_{V}^2)/4M^2} + \e^{s_0/M^2}} \nn\times \bigg\{
		8M^2 m_Q m_{V}^6 f_{V}^T \IA[u_1^3 \hat{\hat{\hat{\mathcal T}}} (\alpha_1,\alpha_3)]
		-4m_Q^3 m_{V}^6 f_{V}^T \IA[u_1^3 \hat{\hat{\hat{\mathcal T}}} (\alpha_1,\alpha_3)]
		\nn -2M^4 m_Q m_{V}^4 f_{V}^T \IA [u_1 \hat{\mathcal T}_3^{(4)} (\alpha_1,\alpha_3) ]
		-2M^4 m_Q m_{V}^4 f_{V}^T \IA[u_1 \hat{\mathcal T}_4^{(4)} (\alpha_1,\alpha_3)]
		\nn - M^6 f_{V} m_{V}^3 \IA[\A]
		- M^6 f_{V} m_{V}^3 \IA[(2u_1-1) \V]
		\bigg\} 
	\end{align}
	where $ u_0 := {M_2^2 \over M_1^2 + M_2^2} = \frac12 $, $ \bar u_0 := 1 - u_0 = \frac12 $, $ u_1 := \frac1{\alpha_3}(-{M^2 \over M_1^2} + 1 - \alpha_1) = \frac1{\alpha_3}(\frac12-\alpha_1) $, we have introduced a shorthand notation for the three-particle DAs as $ \mathcal F(\alpha_1,\alpha_3) := \mathcal F(\alpha_1, \alpha_2 = 1-\alpha_1-\alpha_3, \alpha_3) $, we have defined the alpha-integral operator
	\begin{align}
		\IA[F(\alpha_1, \alpha_3)] & := \int_0^{\bar u_0} \d\alpha_1 \int_{\bar u_0 - \alpha_1} ^{1-\alpha_1} \d\alpha_3\ \frac1{\alpha_3} F(\alpha_1,\alpha_3),
	\end{align}
	and, finally, the hat denotes the following integrations of the DAs:
	\begin{align}
		\hat{\hat f} (u_0) & := \int _0^{u_0} \d v'' \int _0^{v''} \d v'\  f(v'),  \\
		\hat{\mathcal F} (\alpha_1, \alpha_3) &:= \int_0^{\alpha_3} \d\alpha_3' \ \mathcal F (\alpha_1, 1 - \alpha_1 - \alpha_3', \alpha_3'), \\
		\hat{\hat{\hat{\mathcal F}}} (\alpha_1, \alpha_3) &:= \int_0^{\alpha_3} \d\alpha_3''' \int_0^{\alpha_3'''} \d\alpha_3''\int_0^{\alpha_3''} \d\alpha_3'\ \mathcal F (\alpha_1, 1 - \alpha_1 - \alpha_3 ', \alpha_3').
	\end{align}
	\section{Numerical analysis}\label{sec:3}
	In this section, we share the details of our numerical analysis of the LSCR for the strong coupling constants of charmed and bottom mesons $ D^*_{(s)} $, $ D_{(s)1} $, $ B^*_{(s)} $, and $ B_{(s)1} $ with light vector mesons $ \rho $, $ \omega $, $ \phi $, and $ K^* $, where we have used Package X \cite{Patel2015}. The LSCR for the said couplings takes three sets of input parameters. The first and primary set of such parameters are the quark and meson masses and the decay constants of both heavy and light mesons. These are compiled in Table \ref{tab:3.1}. Secondly, there are parameters introduced via the vector meson DAs of different twists. In Appendix \ref{app:A}, we quote only the relevant DAs that appear in our analysis, together with the most up-to-date values of the input parameters for all the light vector mesons taken into account in this work. 
	\begin{table}
		[h]\centering
		\caption{The masses and decay constants of the vector (axial vector) charmed, bottom, and light vector mesons along with the quark masses used in our numerical analysis.}
		\label{tab:3.1}
		\resizebox{\textwidth}{!}{%
		\begin{tabular}
			{|c|c|c|c|c|c|c|c|c|c|c|c|c|c|}
			\hline
			\hline
			Parameter		& Value	& Parameter		& Value					& Parameter		& Value					& Parameter 	& Value					& Parameter 	& Value 					& Parameter		& Value 				& Parameter		& Value \\
			\hline
			$ m_u $ 		& 0		& $ m_\rho $  	& 0.770	\cite{PDG2020}	& $ f_\rho $	& 0.216 \cite{Ball1998}	& $ m_{D^*} $	& 2.010	\cite{PDG2020}	& $ f_{D^*} $	& 0.230 \cite{Wang2008} 	& $ m_{B^*} $	& 5.325 \cite{PDG2020}	& $ f_{B^*} $	& 0.210 \cite{Bazavov2012} \\
			$ m_d $ 		& 0		& $ m_{K^*} $	& 0.892	\cite{PDG2020}	& $ f_\rho^T $ 	& 0.165	\cite{Ball1998}	& $ m_{D_s^*} $	& 2.112 \cite{PDG2020}	& $ f_{D_s^*} $	& 0.266 \cite{Colangelo2005}& $ m_{B_s^*} $	& 5.415 \cite{PDG2020}	& $ f_{B_s^*} $	& 0.251 \cite{Bazavov2012} \\
			$ m_s $ (1 GeV) & 0.137	& $ m_\omega $	& 0.783	\cite{PDG2020}	& $ f_{K^*} $ 	& 0.220	\cite{Ball1998}	& $ m_{D_1} $	& 2.420	\cite{PDG2020}	& $ f_{D_1} $	& 0.219 \cite{Bazavov2012}	& $ m_{B_1} $	& 5.721 \cite{PDG2020}	& $ f_{B_1} $	& 0.335 \cite{Bazavov2012} \\
			$ m_c $ 		& 1.4	& $ m_\phi $	& 1.019	\cite{PDG2020}	& $ f_{K^*}^T $ & 0.185	\cite{Ball1998}	& $ m_{D_{s1}} $& 2.460	\cite{PDG2020}	& $ f_{D_{s1}} $& 0.225 \cite{Thomas2006}	& $ m_{B_{s1}} $& 5.840 \cite{PDG2020}	& $ f_{B_{s1}} $& 0.348 \cite{Bazavov2012} \\
			$ m_b $ 		& 4.8	& 				& 						& $ f_\omega $ 	& 0.187	\cite{Ball1998}	& 				&  						& 				&							&				&						&				& \\
							&		& 				& 						& $ f_\omega^T $& 0.151	\cite{Ball1998} & 				& 						& 				&							&				&						&				& \\
							&		& 				& 						& $ f_\phi $ 	& 0.215	\cite{Ball1998} & 				& 						& 				&							&				&						&				& \\
							&		& 				& 						& $ f_\phi^T $	& 0.186	\cite{Ball1998} & 				& 						& 				&							&				&						&				& \\ 
			\hline
			\hline
		\end{tabular}
		}
	\end{table}
	\par
	Thirdly, the LCSR contains two auxiliary parameters, i.e. the Borel mass parameter, $ M^2 $, and the continuum threshold, $ s_0 $. The coupling constants of the said strong vertices should be independent of the choice of $ M^2 $ and $ s_0 $. This necessitates us to restrict the values of $ M^2 $ and $ s_0 $ to their working regions, which will thus render the LCSR reliable. The lower bound of the Borel mass parameter is obtained by requiring that the contributions from higher-twist terms be well smaller than the leading-twist terms. Its upper bound is determined by considering the fact that the higher-state and continuum contributions should be sufficiently suppressed. These two conditions lead to the following domains of $ M^2 $ that are presented in Table \ref{tab:3.2}. In the meantime, the value of the continuum threshold is obtained by demanding that the two-point sum rules give the mass of the heavy mesons within an accuracy of 10\%. The corresponding values of $ s_0 $ are also presented in Table \ref{tab:3.2}.
	\begin{table}
		[h]\centering
		\caption{The working region of the Borel mass parameter and the continuum threshold for the vertices indicated. Here, $ V_1 = \rho, \omega $.}
		\label{tab:3.2}
		\begin{tabular}
			{|c|c|c||c|c|c|}
			\hline
			\hline
			Vertex & $ M^2 {\rm\ (GeV^2)} $ & $ s_0 {\rm\ (GeV^2)} $ & Vertex & $ M^2 {\rm\ (GeV^2)}$ & $ s_0 {\rm\ (GeV^2)} $\\
			\hline
			$ D^* D^* V_1 $ & $ 3.0 < M^2 < 7.0 $ & $ 6.5 \pm 0.5 $         &  $ B^* B^* V_1 $ & $ 20.0 < M^2 < 24.0 $ & $ 34.0 \pm 1.0 $ \\
			$ D_s^* D_s^* \phi $ & $ 4.0 < M^2 < 8.0 $ & $ 7.0 \pm 0.5 $    &  $ B_s^* B_s^* \phi $ & $ 20.0 < M^2 < 24.0 $ & $ 35.0 \pm 1.0 $ \\
			$ D^* D_s^* K^* $ & $ 5.0 < M^2 < 8.0 $ & $ 6.5 \pm 0.5 $       &  $ B^* B_s^* K^* $ & $ 20.0 < M^2 < 24.0 $ & $ 34.0 \pm 1.0 $ \\
			$ D_1 D_1 V_1 $ & $ 6.0 < M^2 < 10.0 $ & $ 8.5 \pm 1.0 $        &  $ B_1 B_1 V_1 $ & $ 20.0 < M^2 < 24.0 $ & $ 39.0 \pm 1.0 $ \\
			$ D_{s1} D_{s1} \phi $ & $ 6.0 < M^2 < 10.0 $ & $ 9.0 \pm 1.0 $ &  $ B_{s1} B_{s1} \phi $ & $ 20.0 < M^2 < 24.0 $ & $ 40.0 \pm 1.0 $ \\
			$ D_1 D_{s1} K^* $ & $ 6.0 < M^2 < 10.0 $ & $ 9.0 \pm 0.5 $     &  $ B_1 B_{s1} K^* $ & $ 20.0 < M^2 < 24.0 $ & $ 40.0 \pm 1.0 $ \\
			\hline
			\hline
		\end{tabular}
	\end{table}
	\par
\begin{figure}
		[htbp]\centering
		\includegraphics[width=\textwidth]{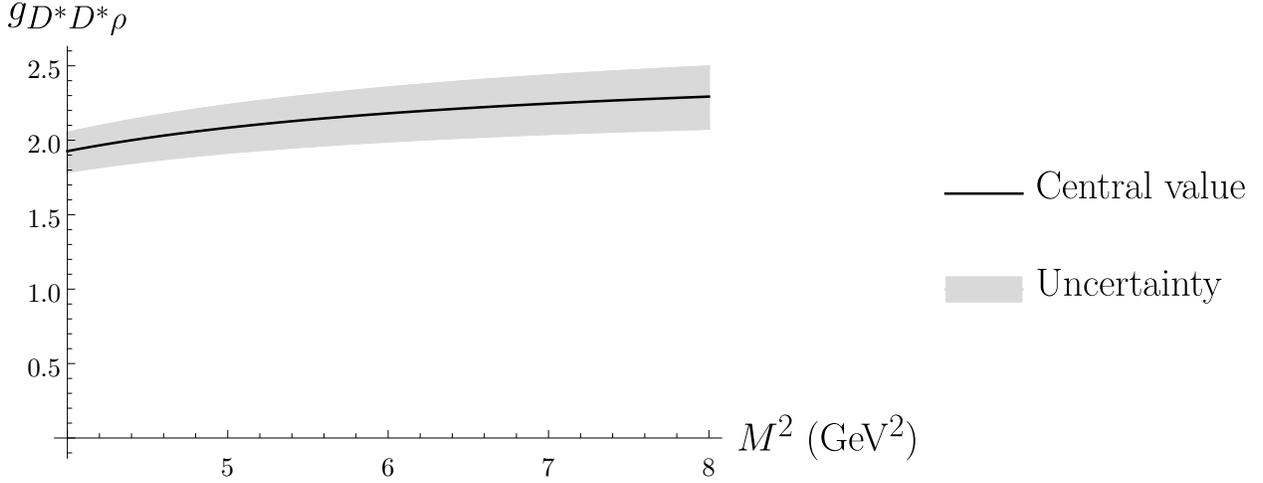}
		\caption{The dependence of the coupling constant of the vertex $ D^*D^*\rho $ on $ M^2 $ at $ s_0 = 6.5 \pm 0.5 {\rm\ GeV^2} $.}
		\label{fig:3.1}
	\end{figure}

\begin{figure}
		[htbp]\centering
		\includegraphics[width=\textwidth]{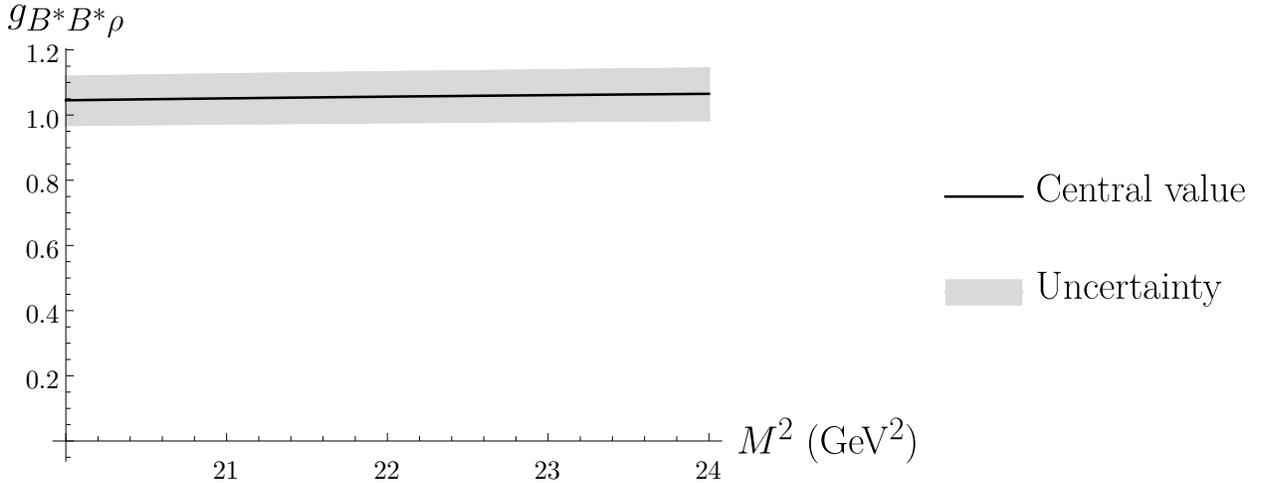}
		\caption{The same as in Fig. \ref{fig:3.1} but for $ B^*B^*\rho $ and at $ s_0 = 34.0 \pm 1.0 {\rm\ GeV^2} $.}
		\label{fig:3.2}
	\end{figure}

 
	Towards the end of our numerical analysis, we found that the leading-twist contributions are around 88\% within the working domains of the Borel mass parameter and at the aforementioned values of the continuum threshold. As an illustration, we present the dependence of the strong coupling constant, $ g $, on the Borel mass parameter, $ M^2 $, for the vertices $ D^*D^*\rho $ and $ B^*B^*\rho $ in Figs. \ref{fig:3.1} and \ref{fig:3.2}, respectively. The coupling constants for $ D^*D^*\rho^0 $, $ D_1D_1\rho^0 $, $ B^*B^*\rho^0 $, and $ B_1B_1\rho^0 $ can be obtained from the corresponding vertices involving the $ \rho^+  $ meson with the help of the isotopic relation.
	\par
	Our estimations for the said coupling constants are presented in Table \ref{tab:3.3.1}. The uncertainties shown in Figs. \ref{fig:3.1} and \ref{fig:3.2} and Table \ref{tab:3.3.1} are due to the variation of the continuum threshold and to the errors in the values of the input parameters.  As we already noted that the $ D^*D^*\rho $ coupling constant within the LCSR method was calculated in \cite{Wang2008}. The difference between our result and that of \cite{Wang2008} lies in the fact that we take into account the contributions of the three-particle DAs, which leads to a small difference between the estimated values of the same coupling constant.
	\def\DstarDstarrho{$ D^*D^*\rho $}
	\def\DstarDstaromega{$ D^*D^*\omega $}
	\def\DsstarDsstarphi{$ D_s^*D_s^*\phi $}
	\def\DstarDsstarKstar{$ D^*D_s^*K^* $}
	\def\DiDirho{$ D_1D_1\rho $}
	\def\DiDiomega{$ D_1D_1\omega $}
	\def\DsiDsiphi{$ D_{s1}D_{s1}\phi $}
	\def\DiDsiKstar{$ D_1D_{s1}K^* $}
	\def\BstarBstarrho{$ B^*B^*\rho $}
	\def\BstarBstaromega{$ B^*B^*\omega $}
	\def\BsstarBsstarphi{$ B_s^*B_s^*\phi $}
	\def\BstarBsstarKstar{$ B^*B_s^*K^* $}
	\def\BiBirho{$ B_1B_1\rho $}
	\def\BiBiomega{$ B_1B_1\omega $}
	\def\BsiBsiphi{$ B_{s1}B_{s1}\phi $}
	\def\BiBsiKstar{$ B_1B_{s1}K^* $}
	\begin{table}
		[htbp]\centering
		\caption{The values of the strong coupling constants of vector (axial vector) $ D $ and $ B $ mesons with light vector mesons. For completeness, the predictions existing in the literature are also presented.}
		\label{tab:3.3.1}
		\resizebox{\textwidth}{!}{%
			\begin{tabular}
				{|c|c|c|c|c|c|c|c|c|c|c|}
				\hline
				\hline
				\multirow{2}{*}{Vertex}	&	\multicolumn{10}{c|}{$ g $}															\\ 
				\cline{2-11}
				&	\multicolumn{2}{c|}{LCSR}			&	3PSR	&	MEM	&	HQCD	&	DSE	&	Lattice	&	PA & Potential & OBE	\\ \hline
				\DstarDstarrho	&	$2.21\pm0.26$	&	$2.6\pm0.7$ \cite{Wang2008}	&	$6.60\pm0.31$ \cite{Bracco2008}, $4.7\pm0.2$ \cite{Bracco2012}	&	2.52 \cite{Oh2001}	&	2.1431 \cite{Bayona2017}	&	10.5, 51.5, 16.8 \cite{ElBennich2017}	&	5.95(56) \cite{Can2013}	&	6.47 \cite{Kim2020} & 3.71 \cite{Lee2009} & 2.6 \cite{Liu2019} 	\\
				\DstarDstaromega	&	$1.70\pm0.23$	&	--	&	--	&	--	&	--	&	--	&	--	&	--	& -- & -- \\
				\DsstarDsstarphi	&	$2.06\pm0.22$	&	--	&	$7.76\pm1.79$ \cite{Khosravi2013}	&	--	&	--	&	--	&	--	&	--	& -- & -- \\
				\DstarDsstarKstar	&	$2.07\pm0.25$	&	--	&	$4.77\pm0.63$ \cite{Janbazi2018}	&	--	&	--	&	--	&	--	&	--	& -- & -- \\
				\DiDirho	&	$2.96\pm0.21$	&	--	&	--	&	--	&	--	&	--	&	--	&	--	& -- & -- \\
				\DiDiomega	&	$2.23\pm0.19$	&	--	&	--	&	--	&	--	&	--	&	--	&	--	& -- & -- \\
				\DsiDsiphi	&	$3.41\pm0.24$	&	--	&	$15.37\pm2.51$ \cite{Khosravi2013}	&	--	&	--	&	--	&	--	&	--	& -- & -- \\
				\DiDsiKstar	&	$3.24\pm0.23$	&	--	&	$4.22\pm0.55$ \cite{Janbazi2018}	&	--	&	--	&	--	&	--	&	--	& -- & -- \\
				\hline
				\BstarBstarrho & $ 1.06\pm0.11 $ & -- & $1.73\pm0.25$ \cite{Cui2012} & 3.71 \cite{Liu2009, Liu2010} & -- & -- & -- & 10.1 \cite{Kim2020} & 3.71 \cite{Lee2009} & 2.6 \cite{Liu2019} \\
				\BstarBstaromega & $ 0.81\pm0.09 $ & -- & -- & -- & -- & -- & -- & -- & -- & -- \\
				\BsstarBsstarphi & $ 1.01\pm0.09 $ & -- & -- & -- & -- & -- & -- & -- & -- & -- \\
				\BstarBsstarKstar & $ 1.05\pm0.11 $ & -- & -- & -- & -- & -- & -- & -- & -- & -- \\
				\BiBirho & $ 0.57\pm0.04 $ & -- & -- & -- & -- & -- & -- & -- & -- & -- \\
				\BiBiomega & $ 0.43\pm0.03 $ & -- & -- & -- & -- & -- & -- & -- & -- & -- \\
				\BsiBsiphi & $ 0.73\pm0.04 $ & -- & -- & -- & -- & -- & -- & -- & -- & -- \\
				\BiBsiKstar & $ 0.66\pm0.04 $ & -- & -- & -- & -- & -- & -- & -- & -- & -- \\
				\hline
				\hline
			\end{tabular}
		}
	\end{table}
	\par
	From Table \ref{tab:3.3.1}, as concerns the couplings involving the $ D $ meson, one can see that our result is in agreement with the result of \cite{Wang2008}, which uses the same framework as ours, within error limits and with that estimated in the holographic QCD; furthermore, it is close to the values predicted by the meson-exchange model and the pole approximation. However, there is a sharp difference between our values and those estimated by the 3PSR, Dyson-Schwinger, and lattice QCD values. For the vertices containing the $ \rho $, $ \phi $, and $ K^* $ mesons, the literature values are nearly two to five times larger than ours for the $ D $ meson sector. For the $ B $ meson sector, comparison of our predictions of the aforementioned couplings with the 3PSR results would be interesting; unfortunately, in the $ B $ meson sector, the said calculations are absent at the present time with an exception for the $ B^*B^*\rho $ vertex, for which the 3PSR method offers a coupling value 2.5 times larger than ours. Nevertheless, there exist strong coupling values for the $ B^*B^*\rho $ vertex obtained within the methods of the pole approximation, meson exchange, potential, and one-boson exchange. Our estimation is two to 10 times smaller than the values determined in the said models. 
	\par
	We would like to further note that our results can be improved by taking into account the $ O(\alpha_s) $ corrections.
	\section{Conclusion}\label{sec:4}
	In this paper, we studied the strong vertices of charmed and bottom mesons $ D_{(s)}^* $, $ D_{(s)1} $, $ B_{(s)}^* $, and $ B_{(s)1} $ with light vector mesons $ \rho $, $ \omega $, $ K^* $, and $ \phi $ within the LCSR method. The said vertices involving the $ D $ mesons are essential in the production of the $ J/\psi $ and $ \phi $ mesons. We have found that our estimation for the coupling of the $ D^*D^*\rho $ vertex agree with the results of \cite{Wang2008}, \cite{Oh2001}, \cite{Bayona2017}, and \cite{Kim2020} but drastically differ from the 3PSR, DSE, and lattice QCD results, whereas our prediction for $ g_{B^*B^*\rho} $ is two to ten times smaller than the values predicted by the pole approximation, the potential model, MEM, and the OBE model.
	\appendix
	\section{Vector meson distribution amplitudes}\label{app:A}
	In this section, we collect the matrix elements $ \bra{V(q,s)} \bar q_1 (x) \Gamma_i q_2(0) \ket 0 $ and \\$ \bra{V(q,s)} \bar q_1 (x) \Gamma_i G_{\lambda\tau}^{(n)} q_2(0) \ket0 $ and the relevant DAs for the light vector mesons that appear in our theoretical analysis, together with the most up-to-date values for the DA parameters involved \cite{Ball1996, Ball1998, Ball1999, Ball2006}. Up to twist-4 accuracy, the said matrix elements are given as follows:
	\begin{align}
		\bra{V(q,s)} \bar q_1 (x) q_2 (0) \ket 0 &= 0,
	\end{align}
	\begin{align}
		\bra{V(q,s)} \bar q_1 (x) \gamma_5 q_2 (0) \ket 0 &= 0,
	\end{align}
	\begin{align}
		\bra{V(q,s)} \bar q_1 (x) \gamma_\mu q_2 (0) \ket 0 &= f_{V} m_{V} \Bigg\{
		\frac{\elx}{\qx} q_\mu \Iu \bb{\phi_2^\parallel(u) + \frac{m_{V}^2 x^2}{16} \phi_4^\parallel(u)}
		\nn + \pp{\el_\mu - q_\mu \frac{\elx}{\qx}} \Iu \phi_3^\perp (u)
		\nn - \frac 12 x_\mu \frac{\elx}{(\qx)^2} m_{V}^2 \Iu \bb{\psi_4^\parallel(u) + \phi_2^\parallel(u) - 2 \phi_3^\perp(u)}
		\Bigg\},
	\end{align}
	\begin{align}
		\bra{V(q,s)} \bar q_1 (x) \I \gamma_\mu \gamma_5 q_2 (0) \ket 0 &= -\frac\I4 \epsilon_{\mu\nu\alpha\beta} {\el}^\nu q^\alpha x^\beta f_{V} m_{V} \Iu \psi_3^\perp(u),
	\end{align}
	\begin{align}
		\bra{V(q,s)} \bar q_1 (x) \frac1{\sqrt2} \sigma_{\mu\nu}& q_2 (0) \ket 0 \nn= -\frac\I{\sqrt2} f_{V}^T \Bigg\{
			(\el_\mu q_\nu - \el_\nu q_\mu) \Iu \bb{\phi_2^\perp(u) + \frac{m_{V}^2x^2}{16} \phi_4^\perp(u)}
			\nn + \frac{\elx}{(\qx)^2} (q_\mu x_\nu - q_\nu x_\mu) \Iu \bb{\phi_3^\parallel(u) - \frac 12 \phi_2^\perp(u)- \frac 12 \psi_4^\perp(u)}
			\nn + \frac 12 (\el_\mu x_\nu - \el_\nu x_\mu) \frac{m_{V}^2}{\qx} \Iu \bb{\psi_4^\perp(u) - \phi_2^\perp(u)}
		\Bigg\},
	\end{align}
	\begin{align}
		\bra{V(q,s)} \bar q_1 (x) G_{\lambda\tau} q_2 (0) \ket 0 &= -\frac\I{g_s} f_{V}^T m_{V} (\el_\lambda q_\tau - \el_\tau q_\lambda) \Ia \mathcal S (\vec\alpha),
	\end{align}
	\begin{align}
		\bra{V(q,s)} \bar q_1 (x) G_{\lambda\tau} \gamma_5 q_2 (0) \ket 0 &= -\frac\I{g_s} f_{V}^T m_{V} \frac12 \epsilon_{\lambda\tau\theta\omega} ({\el}^\theta q^\omega - {\el}^\omega q^\theta) \Ia \tilde{\mathcal S}(\vec\alpha),
	\end{align}
	\begin{align}
		\bra{V(q,s)} \bar q_1 (x) G_{\lambda\tau} \gamma_\mu q_2 (0) \ket 0 &= \frac1{\I g_s} f_{V} m_{V} q_\mu (\el_\lambda q_\tau - \el_\tau q_\lambda) \Ia \mathcal V (\vec\alpha),
	\end{align}
	\begin{align}
		\bra{V(q,s)} \bar q_1 (x) G_{\lambda\tau} \I \gamma_\mu \gamma_5 q_2 (0) \ket 0 &= \frac\I{g_s} f_{V} m_{V} q_\mu \frac12 \epsilon_{\lambda\tau\theta\omega} ({\el}^\theta q^\omega - {\el}^\omega q^\theta) \Ia \mathcal A (\vec\alpha),
	\end{align}
	\begin{align}
		& \bra{V(q,s)} \bar q_1 (x) G_{\lambda\tau} \frac1{\sqrt2} \sigma_{\mu\nu} q_2 (0) \ket 0 \nn = \frac1{\sqrt2 g_s} \Bigg\{
			f_{V}^T m_{V}^2 \frac{\elx}{2\qx} (q_\mu q_\lambda \gper\nu\tau - q_\nu q_\lambda \gper\mu\tau - q_\mu q_\tau \gper\nu\lambda + q_\nu q_\tau \gper\mu\lambda) \Ia \mathcal T (\vec\alpha)
			\nn + f_{V}^T m_{V}^2 (q_\mu \el_\lambda \gper\nu\tau - q_\nu \el_\lambda \gper\mu\tau - q_\mu \el_\tau \gper\nu\lambda + q_\nu \el_\tau \gper\mu\lambda) \Ia \mathcal T_1^{(4)} (\vec\alpha)
			\nn + f_{V}^T m_{V}^2 (q_\lambda \el_\mu \gper\nu\tau - q_\lambda \el_\nu \gper\mu\tau - q_\tau \el_\mu \gper\nu\lambda + q_\tau \el_\nu \gper\mu\lambda) \Ia \mathcal T_2^{(4)} (\vec\alpha)
			\nn + \frac{f_{V}^T m_{V}^2}{\qx} (q_\mu q_\lambda \el_\nu x_\tau - q_\nu q_\lambda \el_\mu x_\tau - q_\mu q_\tau \el_\nu x_\lambda + q_\nu q_\tau \el_\mu x_\lambda) \Ia \mathcal T_3^{(4)} (\vec\alpha)
			\nn + \frac{f_{V}^T m_{V}^2}{\qx} (q_\mu q_\lambda \el_\tau x_\nu - q_\nu q_\lambda \el_\tau x_\mu - q_\mu q_\tau \el_\lambda x_\nu + q_\nu q_\tau \el_\lambda x_\mu) \Ia \mathcal T_4^{(4)} (\vec\alpha)
		\Bigg\}
	\end{align}
	where $ \int \mathcal D  \vec \alpha := \int_0^1 \d \alpha_1 \d \alpha_2 \d\alpha_3 \ \delta(\alpha_1 + \alpha_2 + \alpha_3 - 1) $ and
	\begin{align}
		\gper\mu\nu := g_{\mu\nu} - \frac1{\qx} (b_\mu x_\nu + b_\nu x_\mu)
	\end{align}
	with
	\begin{align}
		b_\mu := q_\mu - \frac{m_{V}^2}{2\qx} x_\mu.
	\end{align}
	Two-particle twist-2 DA:
	\begin{align}
		\phi_2^\parallel(u) &= 6 \bar u u (1+a_1^\parallel C_{1}^{3/2}(\xi)+a_2^\parallel C_{2}^{3/2}(\xi)).
	\end{align}
	Two-particle twist-3 DA:
	\begin{align}
		\phi_3^\perp (u) &= (3 a_1^\parallel \xi^3)/2+3/4 (1+\xi^2)+(5 \kappa_3^\parallel-(15 \lambda_3^\parallel)/16+(15 \tilde\lambda_3^\parallel)/8) \xi (-3+5 \xi^2) \nn +((9 a_2^\parallel)/112+(15 \omega_3^\parallel)/32-(15 \tilde\omega_3^\parallel)/64) (3-30 \xi^2+35 \xi^4)+(-1+3 \xi^2) \nn\times ((3 a_2^\parallel)/7+5 \zeta_3^\parallel)-(1/(2 f_{V} m_{V}))3 f_{V}^T (m_{q_1}-m_{q_2}) (2 \xi+2 a_2^\perp \xi (11-20 \bar u u) \nn +9 a_1^\perp (1-2 \bar u u)+(1+3 a_1^\perp+6 a_2^\perp) \ln(\bar u)-(1-3 a_1^\perp+6 a_2^\perp) \ln(u)) \nn +(1/(2 f_{V} m_{V}))3 f_{V}^T (m_{q_1}+m_{q_2}) (2+9 a_1^\perp \xi+2 a_2^\perp (11-30 \bar u u)+(1+3 a_1^\perp \nn +6 a_2^\perp) \ln(\bar u)+(1-3 a_1^\perp+6 a_2^\perp) \ln(u)).
	\end{align}
	Two-particle twist-4 DAs:
	\begin{align}
		\psi_4^\parallel(u) &= 1+(1/(f_{V} m_{V}))6 f_{V}^T (m_{q_1}-m_{q_2}) (\xi+1/2 a_1^\perp (-1+3 \xi^2)+5/2 \kappa_3^\perp (-1 \nn +3 \xi^2)+1/2 a_2^\perp \xi (-3+5 \xi^2)+5/6 \omega_3^\perp \xi (-3+5 \xi^2)-1/16 \lambda_3^\perp (3-30 \xi^2 \nn +35 \xi^4))+((9 a_1^\parallel)/5+12 \kappa_4^\parallel) C_{1}^{1/2}(\xi)+(-1-(2 a_2^\parallel)/7+(40 \zeta_3^\parallel)/3) C_{2}^{1/2}(\xi) \nn -(20 \zeta_4^\parallel C_{2}^{1/2}(\xi))/3+(-((9 a_1^\parallel)/5)-(20 \kappa_3^\parallel)/3-(16 \kappa_4^\parallel)/3) C_{3}^{1/2}(\xi)+(10 \theta_1^\parallel \nn -5 \theta_2^\parallel) C_{3}^{1/2}(\xi)+(-((27 a_2^\parallel)/28)-(15 \omega_3^\parallel)/8-(15 \tilde\omega_3^\parallel)/16+(5 \zeta_3^\parallel)/4) C_{4}^{1/2}(\xi), \\
		\phi_4^\parallel(u) &= (1/(f_{V} m_{V}))f_{V}^T (m_{q_1}-m_{q_2}) ((-23-54 a_1^\perp-108 a_2^\perp+5 u^2) \ln(\bar u)-(-23 \nn +54 a_1^\perp-108 a_2^\perp+5 (\bar u)^2) \ln(u))+(1/(f_{V} m_{V}))24 f_{V}^T (m_{q_1}+m_{q_2}) ((1+3 a_1^\perp \nn +6 a_2^\perp) (\bar u)^2 \ln(\bar u)+(1-3 a_1^\perp+6 a_2^\perp) u^2 \ln(u))+4 (a_1^\parallel-(40 \kappa_3^\parallel)/3) (1/8 (11 \nn -3 \xi^2)-(2-\bar u) (\bar u)^3 \ln(\bar u)+(2-u) u^3 \ln(u))+80 \psi_2^\parallel (1/8 (11-3 \xi^2) \nn -(2-\bar u) (\bar u)^3 \ln(\bar u)+(2-u) u^3 \ln(u))-80 \tilde\omega_4^\parallel (1/8 \bar u (21-13 \xi^2) u \nn +(\bar u)^3 (10-15 \bar u+6 (\bar u)^2) \ln(\bar u)+u^3 (10-15 u+6 u^2) \ln(u))+2 (-2 a_2^\parallel \nn +3 \omega_3^\parallel-(14 \zeta_3^\parallel)/3) (1/8 \bar u (21-13 \xi^2) u+(\bar u)^3 (10-15 \bar u+6 (\bar u)^2) \ln(\bar u) \nn +u^3 (10-15 u+6 u^2) \ln(u))+30 (\bar u)^2 u^2 ((20 \zeta_4^\parallel)/9+(-((8 \theta_1^\parallel)/15) \nn +(2 \theta_2^\parallel)/3) C_{1}^{5/2}(\xi))+30 (\bar u)^2 u^2 (4/5 (1+(a_2^\parallel)/21+(10 \zeta_3^\parallel)/9)+((17 a_1^\parallel)/50 \nn -(\lambda_3^\parallel)/5+(2 \tilde\lambda_3^\parallel)/5) C_{1}^{5/2}(\xi)+1/10 ((9 a_2^\parallel)/7+(7 \omega_3^\parallel)/6-(3 \tilde\omega_3^\parallel)/4 \nn +(\zeta_3^\parallel)/9) C_{2}^{5/2}(\xi))+(1/(f_{V} m_{V}))6 \bar u f_{V}^T (m_{q_1}+m_{q_2}) u (2 (3+16 a_2^\perp) \nn +10/3 (-a_1^\perp+\kappa_3^\perp) C_{1}^{3/2}(\xi)+(-a_2^\perp+(5 \omega_3^\perp)/9) C_{2}^{3/2}(\xi)-(\lambda_3^\perp C_{3}^{3/2}(\xi))/10) \nn +(1/(f_{V} m_{V}))6 \bar u f_{V}^T (m_{q_1}-m_{q_2}) u ((-((82 a_1^\perp)/5)-10 \kappa_3^\perp) C_{1}^{3/2}(\xi) \nn +20 (10/189+(a_2^\perp)/3-(\omega_3^\perp)/21) C_{2}^{3/2}(\xi)+((2 a_1^\perp)/5+(7 \lambda_3^\perp)/54) C_{3}^{3/2}(\xi) \nn +(-(2/315)+(a_2^\perp)/5-(\omega_3^\perp)/21) C_{4}^{3/2}(\xi)+(2 \lambda_3^\perp C_{5}^{3/2}(\xi))/135).
	\end{align}
	Three-particle twist-3 DAs:
	\begin{align}
		\mathcal V (\vec \alpha) = \mathcal V (\alpha_1, \alpha_2, \alpha_3) &= 360 \alpha_1 (1-\alpha_1-\alpha_3) \alpha_3^2 (\kappa_3^\parallel+1/2 (-3+7 \alpha_3) \lambda_3^\parallel+(-1+2 \alpha_1+\alpha_3) \omega_3^\parallel), \\ 
		\mathcal A (\vec \alpha) = \mathcal A (\alpha_1, \alpha_2, \alpha_3) &= 360 \alpha_1 (1-\alpha_1-\alpha_3) \alpha_3^2 ((-1+2 \alpha_1+\alpha_3) \tilde\lambda_3^\parallel+1/2 (-3+7 \alpha_3) \tilde\omega_3^\parallel+\zeta_3^\parallel), \\
		\mathcal T (\vec \alpha) = \mathcal T (\alpha_1, \alpha_2, \alpha_3) &= 360 \alpha_1 (1-\alpha_1-\alpha_3) \alpha_3^2 (\kappa_3^\perp+1/2 (-3+7 \alpha_3) \lambda_3^\perp+(-1+2 \alpha_1+\alpha_3) \omega_3^\perp).
	\end{align}
	Three-particle twist-4 DAs:
	\begin{align}
		\mathcal T_3^{(4)} (\vec \alpha) = \mathcal T_3^{(4)} (\alpha_1, \alpha_2, \alpha_3) &= -120 \alpha_1 (1-\alpha_1-\alpha_3) \alpha_3 (\tilde\phi_0^\perp+(-1+2 \alpha_1+\alpha_3) \tilde\phi_1^\perp+(-1+3 \alpha_3) \tilde\phi_2^\perp), \\
		\mathcal T_4^{(4)} (\vec \alpha) = \mathcal T_4^{(4)} (\alpha_1, \alpha_2, \alpha_3) &= 30 \alpha_3^2 (-((-1+2 \alpha_1+\alpha_3) (\psi_0^\perp+\alpha_3 \psi_1^\perp+1/2 (-3+5 \alpha_3) \psi_2^\perp)) \nn +(1-\alpha_3) \theta_0^\perp+(-6 \alpha_1 (1-\alpha_1-\alpha_3)+(1-\alpha_3) \alpha_3) \theta_1^\perp+(-(3/2) ((\alpha_1)^2 \nn +(1-\alpha_1-\alpha_3)^2)+(1-\alpha_3) \alpha_3) \theta_2^\perp).
	\end{align}
	Numerical values of the parameters that appear in the DAs are compiled in Table \ref{tab:A.1}.
	\begin{table}
		\centering
		\caption{All the values are at the energy scale of $ \mu = 1 {\rm\ GeV} $. The accuracy of these parameters are 30--50\%.}
		\label{tab:A.1}
		\resizebox{\textwidth}{!}{%
			\begin{tabular}{*{20}{|c}|}%
			\hline
			\hline
			& $ a_1^\parallel $ & $ a_1^\perp $ & $ a_2^\parallel $ & $ a_2^\perp $ & $ \zeta_3^\parallel $ & $ \tilde\lambda_3^\parallel $ & $ \tilde\omega_3^\parallel $ & $ \kappa_3^\parallel $ & $ \omega_3^\parallel $ & $ \lambda_3^\parallel $ & $ \kappa_3^\perp $ & $ \omega_3^\perp $ & $ \lambda_3^\perp $ & $ \zeta_4^\parallel $ & $ \tilde\omega_4^\parallel $ & $ \zeta_4^\perp $ & $ \tilde \zeta _4 ^\perp $ & $ \kappa_4^\parallel $ & $ \kappa_4^\perp $ \\
			\hline
			$ \rho^+ $ & 0 & 0 & 0.15 & 0.14 & 0.030 & 0 & --0.09 & 0 & 0.15 & 0 & 0 & 0.55 & 0 & 0.07 & --0.03 & --0.03 & --0.08 & 0 & 0 \\
			$ K^* $ & 0.03 & 0.04 & 0.11 & 0.10 & 0.023 & 0.035 & --0.07 & 0 & 0.1 & --0.008 & 0.003 & 0.03 & --0.025 & 0.02 & --0.02 & --0.01 & --0.05 & --0.025 & 0.013 \\
			$ \omega $ & 0 & 0 & 0.15 & 0.14 & 0.030 & 0 & --0.09 & 0 & 0.15 & 0 & 0 & 0.55 & 0 & 0.07 & --0.03 & --0.03 & --0.08 & 0 & 0 \\
			$ \phi $ & 0 & 0 & 0.18 & 0.14 & 0.024 & 0 & --0.045 & 0 & 0.09 & 0 & 0 & 0.20 & 0 & 0 & --0.02 & --0.01 & --0.03 & 0 & 0 \\
			\hline
		\end{tabular}%
		}
	\end{table}
	\section{Important integrals}\label{app:B}
	In this section, we share the results of various integrals that appear in the theoretical side of the correlation function. We grouped the integrals into two: those coming from the terms that do not involve the gluon and those that contain it. From the gluon terms, there will come three-particle DAs, say $ \mathcal F (\alpha_1, \alpha_2, \alpha_3) =: \mathcal F(\vec \alpha) $ where $ \mathcal F = \mathcal S, \tilde{\mathcal S}, \mathcal V, \mathcal A, \mathcal T, \mathcal T_k^{(4)} $ with $ k = 1,2,3,4 $.
	\par
	We do the terms without the gluon first. Let $ K_n := K_n(m_Q \sqrt{-x^2})/\pp{\sqrt{-x^2}}^n $ and
	\begin{align}
		I^n & := \int_0^1 \d u \int \D4x\ \e^{\I (p+\bar u q)x} K_n f(u), \\
		I^n_\mu & := \int_0^1 \d u \int \D4x\ \e^{\I (p+\bar u q)x} x_\mu K_n f(u), \\
		I^n_{\mu\nu} & := \int_0^1 \d u \int \D4x\ \e^{\I (p+\bar u q)x} x_\mu x_\nu K_n f(u), \\
		I^n_{\mu\nu\lambda} & := \int_0^1 \d u \int \D4x\ \e^{\I (p+\bar u q)x} x_\mu x_\nu x_\lambda K_n f(u), \\
		I^n_2 & := \int_0^1 \d u \int \D4x\ \e^{\I (p+\bar u q)x} x^2 K_n f(u), \\
		I^n_{2\mu} & := \int_0^1 \d u \int \D4x\ \e^{\I (p+\bar u q)x} x^2 x_\mu K_n f(u), \\
		I^n_{2\mu\nu} & := \int_0^1 \d u \int \D4x\ \e^{\I (p+\bar u q)x} x^2 x_\mu x_\nu K_n f(u) .
	\end{align}
	After a double Borel transformation over $ -p^2 $ and $ -(p+q)^2 $ and performing the continuum subtraction via the replacement $ \e^{-m_Q^2/M^2 - m_V^2/(M_1^2 + M_2^2)} \to \e^{-m_Q^2/M^2 - m_V^2/(M_1^2 + M_2^2)} - \e^{-s_0/M^2} $ as described in \cite{Belyaev1995}, one has
	\begin{align}
		I^n & \to \I \frac{2^{3-n}\pi^2}{m_Q^n} M^{2n} \bb{\e^{-m_Q^2/M^2 - m_{V}^2/(M_1^2 + M_2^2)} - \e^{-s_0/M^2}} f(u_0),\\
		I^n_\mu & \to -\frac{2\I (p+\bar u_0 q)_\mu}{M^2} I^n, \\
		I^n_{\mu\nu} & \to - \frac{2[g_{\mu\nu} M^2 + 2(p+ \bar u_0 q)_\mu (p + \bar u_0 q)_\nu]}{M^4} I^n, \\
		I^n_{\mu\nu\lambda} &\to \frac{4\I}{M^6} \{
			[M^2 g_{\mu\nu} (p+ \bar u_0 q)_\lambda 
			+ M^2 g_{\nu\lambda} (p+ \bar u_0 q)_\mu 
			+ M^2 g_{\mu\lambda} (p+ \bar u_0 q)_\nu 
			\nn + 2 (p+ \bar u_0 q)_\mu (p+ \bar u_0 q)_\nu (p+ \bar u_0 q)_\lambda
		\} I^n, \\
		I_2^n & \to -\I \frac{2^{5-n}\pi^2}{m_Q^n} M^{2n-4} [m_Q^2 + (n-1) M^2] \bb{\e^{-m_Q^2/M^2 - m_{V}^2/(M_1^2 + M_2^2)} - \e^{-s_0/M^2}} f(u_0),\\
		I^n_{2\mu} & \to -\frac{2\I(p+\bar u_0q)_\mu}{M^2} I_2^n, \\
		I^n_{2\mu\nu} & \to -\frac{2[M^2 g_{\mu\nu} + 2 (p+\bar u_0q)_\mu (p+\bar u_0q)_\nu]}{M^4} I_2^n
	\end{align}
	where $ u_0 := M_2^2/(M_1^2+M_2^2) $, $ \bar u_0 := 1-u_0 = M^2/M_1^2+M_2^2 $, and $ M $ is defined via $ (M^2)^{-1} = (M_1^2)^{-1} + (M_2^2)^{-1} $. For nearly-equal-mass mesons, one can take $ M_1 = M_2 $ and hence $ u_0 = \bar u_0 = 1/2 $.
	\par
	Next, let's do the terms with the gluon. Let
	\begin{align}
		J^n & := \int_0^1 \d u \int \D4x \int \mathcal D \vec \alpha\ \e^{\I [p+(\alpha_1 + u \alpha_3) q]x} K_n f(u) F(\vec \alpha), \\
		J^n_\mu & := \int_0^1 \d u \int \D4x \int \mathcal D \vec \alpha\ \e^{\I [p+(\alpha_1 + u \alpha_3) q]x} x_\mu K_n f(u) F(\vec \alpha), \\
		J^n_{\mu\nu} & := \int_0^1 \d u \int \D4x \int \mathcal D \vec \alpha\ \e^{\I [p+(\alpha_1 + u \alpha_3) q]x} x_\mu x_\nu K_n f(u) F(\vec \alpha), \\
		J^n_{\mu\nu\lambda} & := \int_0^1 \d u \int \D4x \int \mathcal D \vec \alpha\ \e^{\I [p+(\alpha_1 + u \alpha_3) q]x} x_\mu x_\nu x_\lambda K_n f(u) F(\vec \alpha), \\
		J^n_2 & := \int_0^1 \d u \int \D4x \int \mathcal D \vec \alpha\ \e^{\I [p+(\alpha_1 + u \alpha_3) q]x} x^2 K_n f(u) F(\vec \alpha), \\
		J^n_{2\mu} & := \int_0^1 \d u \int \D4x \int \mathcal D \vec \alpha\ \e^{\I [p+(\alpha_1 + u \alpha_3) q]x} x^2 x_\mu K_n f(u) F(\vec \alpha), \\
		J^n_{2\mu\nu} & := \int_0^1 \d u \int \D4x \int \mathcal D \vec \alpha\ \e^{\I [p+(\alpha_1 + u \alpha_3) q]x} x^2 x_\mu x_\nu K_n f(u) F(\vec \alpha).
	\end{align}
	where $ \int \mathcal D \vec\alpha := \int_0^1 \d\alpha_1 \d\alpha_2 \d\alpha_3 \ \delta(\alpha_1 + \alpha_2 + \alpha_3 - 1) $. The results are as follows:
	\begin{align}
		J^n & \to \I \frac{2^{3-n}\pi^2}{m_Q^n} M^{2n} \bb{\e^{-m_Q^2/M^2 - m_{V}^2/(M_1^2 + M_2^2)} - \e^{-s_0/M^2}} \nn\times \int_0^{\bar u_0} \d\alpha_1 \int_{\bar u_0 - \alpha_1} ^{1-\alpha_1} \d\alpha_3\ \frac 1{\alpha_3} F(\alpha_1, 1 - \alpha_1 - \alpha_3, \alpha_3) f(u_1), \\
		J^n_\mu & \to -\frac{2\I (p+ \bar u_0 q)_\mu}{M^2} J^n, \\
		J^n_{\mu\nu} & \to - \frac{2[g_{\mu\nu} M^2 + 2(p+ \bar u_0 q)_\mu (p + \bar u_0 q)_\nu]}{M^4} J^n, \\
		J^n_{\mu\nu\lambda} &\to \frac{4\I}{M^6} \{
		[M^2 g_{\mu\nu} (p+ \bar u_0 q)_\lambda 
		+ M^2 g_{\nu\lambda} (p+ \bar u_0 q)_\mu 
		+ M^2 g_{\mu\lambda} (p+ \bar u_0 q)_\nu ] \nn + 2 (p+ \bar u_0 q)_\mu (p+ \bar u_0 q)_\nu (p+ \bar u_0 q)_\lambda
		\} J^n, \\
		J_2^n & \to -\I \frac{2^{5-n}\pi^2}{m_Q^n} M^{2n-4} [m_Q^2 + (n-1) M^2] \bb{\e^{-m_Q^2/M^2 - m_{V}^2/(M_1^2 + M_2^2)} - \e^{-s_0/M^2}} \nn\times \int_0^{\bar u_0} \d\alpha_1 \int_{\bar u_0 - \alpha_1} ^{1-\alpha_1} \d\alpha_3\ \frac 1{\alpha_3} F(\alpha_1, 1 - \alpha_1 - \alpha_3, \alpha_3) f(u_1), \\
		J^n_{2\mu} & \to -\frac{2\I (p+ \bar u_0 q)_\mu}{M^2} J_2^n, \\
		J^n_{2\mu\nu} & \to - \frac{2[g_{\mu\nu} M^2 + 2(p+ \bar u_0 q)_\mu (p + \bar u_0 q)_\nu]}{M^4} J_2^n, \\
		J^n_{2\mu\nu\lambda} &\to \frac{4\I}{M^6} \{
		[M^2 g_{\mu\nu} (p+ \bar u_0 q)_\lambda 
		+ M^2 g_{\nu\lambda} (p+ \bar u_0 q)_\mu 
		+ M^2 g_{\mu\lambda} (p+ \bar u_0 q)_\nu ] \nn + 2 (p+ \bar u_0 q)_\mu (p+ \bar u_0 q)_\nu (p+ \bar u_0 q)_\lambda
		\} J_2^n 
	\end{align}
	where $ u_1 := \frac1{\alpha_3} (1-{M^2 \over M_1^2} - \alpha_1) $.
	\bibliography{refs}
\end{document}